\documentclass{article}
\usepackage{spconf,amsmath,graphicx,tikz,amssymb}

\title{Stable Optimization for Large Vision Model Based \\ Deep Image Prior in Cone-Beam CT Reconstruction}
\name{Minghui Wu$^1$, Yangdi Xu$^1$, Yingying Xu$^1$, Guangwei Wu$^1$, Qingqing Chen$^2$, Hongxiang Lin$^{1,}$\sthanks{Corresponding author: Hongxiang Lin (Email: harryhxlin@gmail.com).
This work was supported in part by the National Natural Science Foundation of China under Grant No. 12301546 and Zhejiang Provincial Natural Science Foundation of China under Grant No. LQ24F010007.}}
\address{$^1$ Zhejiang Lab; $^2$ Sir Run Run Shaw Hospital, Zhejiang University College of Medicine}

\begin{document}
%
\maketitle
\begin{abstract}
    Large Vision Model (LVM) has recently demonstrated great potential for medical imaging tasks, potentially enabling image enhancement for sparse-view Cone-Beam Computed Tomography (CBCT), despite requiring a substantial amount of data for training. Meanwhile, Deep Image Prior (DIP) effectively guides an untrained neural network to generate high-quality CBCT images without any training data. However, the original DIP method relies on a well-defined forward model and a large-capacity backbone network, which is notoriously difficult to converge. In this paper, we propose a stable optimization method for the forward-model-free, LVM-based DIP model for sparse-view CBCT. Our approach consists of two main characteristics: (1) multi-scale perceptual loss (MSPL) which measures the similarity of perceptual features between the reference and output images at multiple resolutions without the need for any forward model, and (2) a reweighting mechanism that stabilizes the iteration trajectory of MSPL. One shot optimization is used to simultaneously and stably reweight MSPL and optimize LVM. We evaluate our approach on two publicly available datasets: SPARE and Walnut. The results show significant improvements in both image quality metrics and visualization that demonstrates reduced streak artifacts. The source code is available upon request.
\end{abstract}

\section{Introduction}
\label{sec:intro}
Cone-Beam Computed Tomography (CBCT) has become increasingly prevalent in clinical settings, particularly for imaging applications in dentistry, maxillofacial surgery, and lung examination. Compared to conventional 2D CT methods, CBCT can obtain 3D tomographic images at an equivalent radiation dose but with faster data acquisition process. Classical model-based CBCT reconstruction methods, such as Feldkamp-Davis-Kress (FDK)~\cite{feldkamp1984practical} and Simultaneous Iterative Reconstruction Technique (SIRT)~\cite{kak2001principles}, are commonly used in practice but may degrade the reconstructed images by introducing streak artifacts, particularly for sparse-view acquisition of measurements. In contrast, recent deep learning technology seamlessly integrates neural networks with the model-based methods, rapidly achieving state-of-the-art performance but often requiring massive data for training~\cite{2019Augmentation}. Some research indicates that such model trained on synthetic data may be significantly prone to hallucination which introduces false positive structure into the image anatomy~\cite{Bhadra2021}. 


Deep image prior (DIP) utilizes a generative neural network to produce and restore high-quality images in a neural form~\cite{Ulyanov2020}. DIP has successfully developed for CT~\cite{2019Low}, which does not require any training process on large datasets but optimizes the residual of CT forward equation based on a single measurement; such idea can be generalized to CBCT. Meanwhile, recent advancement of large vision model (LVM)~\cite{wang2023mathrmsammed} including Vision transformer (ViT)~\cite{dosovitskiy2020image} and Segment Anything Model (SAM)~\cite{kirillov2023segment} in medical imaging further motivates us to apply LVM to DIP, whereas such models having transformer modules are notoriously hard to optimize~\cite{Chen2021,Lin2022}.

Therefore, we propose an unsupervised forward-model-free LVM-based DIP for CBCT reconstruction, whose model serves as the CBCT inverse solver without the need of large number of measurements, ground-truth images, or geometric information. We attain stable optimization through two main characteristics: the multi-scale perceptual loss (MSPL) and the reweighting mechanism. First, MSPL measures how perceptually similar the two images are at different layers of a pre-trained network that corresponds to multiple resolution scales~\cite{Johnson2016,zhang2018unreasonable}. MSPL can preserve the feature-level fidelity of generative high-quality image and sparse-view reference image. Second, the reweighting mechanism sets all the weights in MSPL to be learnable and adopts one shot optimization method to simultaneously update all those learnable parameters within both loss function and neural network.

Our contributions are as follows: (1) Deriving the first DIP method with an LVM backbone for 3D CBCT using only one measurement data without any ground-truth image; (2) Stabilizing the optimization process of our proposed DIP based approach that is incorporated with an LVM backbone; (3) Automatically reweighting MSPL to boost performance of our approach for 3D CBCT reconstruction.

\section{Method}
\label{sec:method}

\subsection{Model}
\label{sec:untrain}

\begin{figure}[t]
\centering
\includegraphics[width=.5\textwidth]{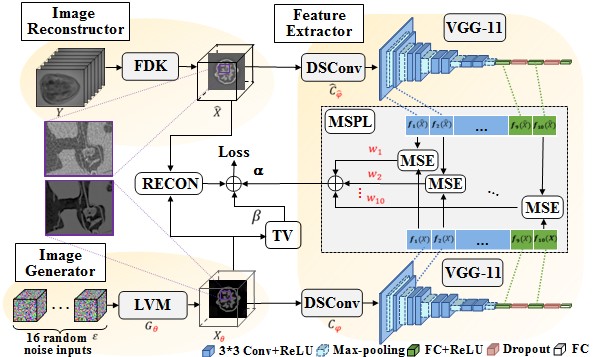}
\caption{The framework of the proposed forward-model-free, LVM-based DIP with MSPL for sparse-view CBCT reconstruction. Texts with red highlight the learnable variables including loss and network weights.}
\label{fig:1}
\end{figure}

We first formulate the LVM-based DIP model for sparse-view 3D CBCT image reconstruction. As illustrated in Fig.~\ref{fig:1}, the model mainly consists of the following three modules:

\textbf{Image Reconstructor.} This module uses the FDK algorithm~\cite{feldkamp1984practical} to obtain a reference image $\hat{X}\in \mathbb{R}^{N_1\times N_2\times N_3}$, given the sparse-view measurement (projection data) $Y$.

\textbf{Image Generator.} This adapts to an LVM architecture $G_\theta$ parameterized by $\theta$, which maps a 4D random noise volume with $16$ channels $\varepsilon \in \mathbb{R}^{16\times N_1\times N_2\times N_3}$ to the generating CBCT image $X_{\theta}\in \mathbb{R}^{N_1\times N_2\times N_3}$. We adopt a modified UNETR architecture~\cite{hatamizadeh2022unetr} as an example LVM module. The UNETR features a U-Net~\cite{cciccek20163d} architecture that consists of token (or 3D patch) extraction, linear projection with positional embedding, Transformer encoding and DNNs decoding with U-Net style skip connection.

\textbf{Feature Extractor.} This consists of the Dimension Shrinkage Convolution (DSConv) operation and a pre-trained encoder network, e.g. VGG-11 pre-trained on the ImageNet~\cite{Simonyan2015}. The DSConv operation $C_\varphi:\mathbb{R}^{N_1\times N_2\times N_3}\rightarrow \mathbb{R}^{3\times N_2\times N_3}$, parameterized by $\varphi$, is a two-dimensional convolution whose channel at the first dimension reduces to the number of the input channels ($3$ here) of VGG-11.

\subsection{Loss Function}
We aim to optimize the above network model by minimizing a loss function in terms of learnable weights. The loss function consists of three parts: Reconstruction (RECON) Loss, MSPL, and the total variation (TV) penalty term. 

First, unlike the original DIP~\cite{Ulyanov2020}, we do not integrate the forward model into RECON but directly minimize the mean squared error (MSE), denoted by $\|\cdot\|_2^2$, between the generative image $X_\theta$ and the sparse-view reference image $\hat{X}$ as follows:
\begin{equation}\label{eq:recon}
    L_{1}(\theta) = \Vert\hat{X}-X_\theta\Vert_2^{2},
\end{equation}
where $\theta$ denotes the network parameter in Image Generator.

Second, our hypothesis is that minimizing MSPL can complement the lack of forward model. MSPL measures perceptual feature difference at multiple resolution scales. In specific, it is weighted sum of a mixture of all the activation maps in the pre-trained VGG-11 network, which obtain the expression for MSPL:
\begin{equation}
L_{2}(\theta, \varphi, \hat{\varphi}, \mathbf{w})=\sum_{i=1}^K w_{i}\Vert f_{i}(\hat{C}_{\hat{\varphi}}(\hat{X}))-f_{i}({C}_{{\varphi}}(X_\theta))\Vert_{2}^{2},
\label{eq:MSPL}
\end{equation}
where $\mathbf{w}=(w_1, \cdots, w_K)$ is the weights of MSPL of $K$ resolution scales, $\varphi$ and $\hat{\varphi}$ are the network parameters of the reconstruction DSConv $C_\varphi$ and the generation DSConv $\hat{C}_{\hat{\varphi}}$, respectively, and $f_i$ is the intermediate output of VGG-11 at the $i$th Rectified Linear Unit (ReLU) layer. 

Lastly, together with Equations (\ref{eq:recon}-\ref{eq:MSPL}) and the TV penalty term built in~\cite{rudin1992}, we have the total loss function $L$:
\begin{equation}
L(\theta, \varphi, \hat{\varphi}, \mathbf{w})=L_{1}(\theta)+\alpha L_{2}(\theta, \varphi, \hat{\varphi}, \mathbf{w})+\beta TV(\theta),
\label{eq:multiLoss}
\end{equation}
where $\alpha, \beta$ is non-negative coefficients. During the optimization process, we freeze the network parameters of the two VGG-11 feature extractors.

\subsection{Reweighting Mechanism for Stable Optimization} 
\label{sec:reweight}
\begin{figure*}[t]
    \centering
\includegraphics[width=0.9\textwidth]{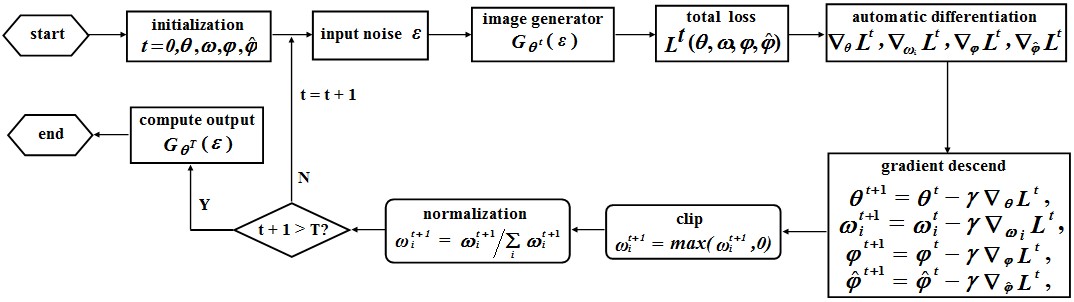}
    \caption{The flow chart of the adaptive reweighting one shot optimization algorithm.} 
    \label{fig:4}
\end{figure*}

Our reweighting mechanism is inspired by the one shot optimization in~\cite{guenther2016simultaneous}, which can simultaneously update all learnable parameters during one iteration step. In our case, a successful choice of MSPL weights $\mathbf{w}$ in Eq.~\eqref{eq:MSPL} ensures the stable optimization of our proposed method. Instead of using a group of fixed heuristic values, we propose a novel adaptive reweighting one shot optimization that takes $\mathbf{w}$ as differentiable variables and subsequently reweight them along with all the network parameters. As demonstrated in 
Fig.~\ref{fig:4}, we use automatic differentiation (AD) to compute the gradient of the total loss in Eq.~\eqref{eq:multiLoss} at the iteration step $t$, and use a gradient descend (GD) method (the ADAM optimizer~\cite{kingma2014adam} in our experiment) to update those parameters with the learning rate $\gamma$. It is noting that the adaptive weight component $w_{i}$ may occasionally become negative, which leads to performance degradation of optimization due to the negative loss values. To solve this, we employ the clip/gradient clip and the subsequent normalization operations on $\mathbf{w}$ during each iteration step to avoid the negative loss phenomenon and further stabilize the optimization process.

\section{Experiment and Results}
\label{sec:results}

We conducted our study on the SPArse-view REconstruction (SPARE) Challenge dataset~\cite{Shieh2019} and the Walnut dataset~\cite{DerSarkissian2019} which are publicly available. We used the FDK algorithm to construct both the ground-truth (GT) volumes from full-view measurements and the reference volumes from sparse-view measurements. All the experiments were conducted on Pytorch 1.9 and ran on a single Nvidia Tesla V100 GPU with $32$ gigabytes memory. We used the FDK and the SIRT algorithms to obtain sparse-view reconstructed images for comparison. The iteration number for SIRT was set to $200$. The ADAM optimizer was utilized with an initial learning rate of $10^{-3}$ and decay of $10^{-5}$. The regularization parameter $\alpha$ was $1$. The inputs random noise volumes were randomly drawn from the 3D standard Gaussian distribution. All backbone networks were initialized by the Glorot normal initializer. For the UNETR backbone, the patch size was set to $16\times16\times16$. To unify input feature dimensions, the patch-based 3D U-Net baseline chose the same input shape. All DIP-based methods were terminated at $2000$th iteration from our best practice. Image quality was quantitatively evaluated by peak signal-to-noise ratio (PSNR) and structural similarity index (SSIM) between the reconstruction result and the GT image.

\begin{figure*}[ht]
\centering
\includegraphics[width=0.9\textwidth]{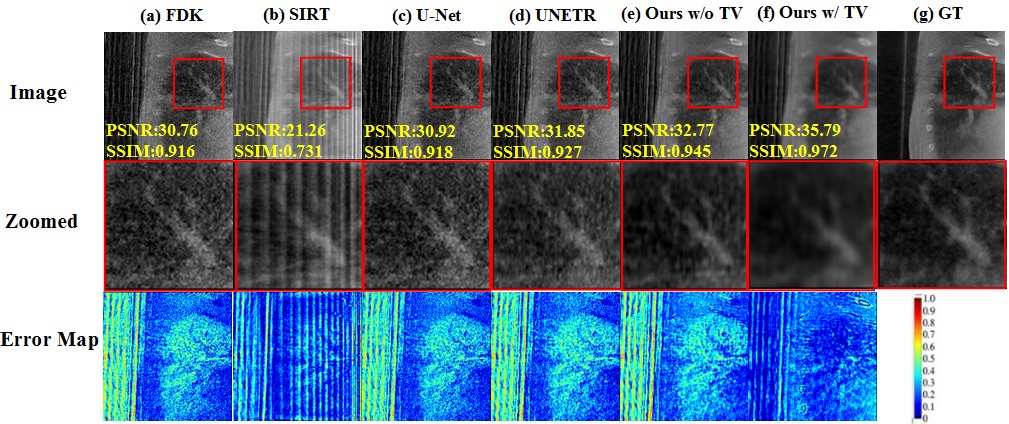}
\caption{ Visualization of the sagittal views by (a) FDK, (b) SIRT, (c) DIP with the 3D U-Net backbone, (d) DIP with the UNETR backbone, (e) Our proposed approach w/o TV, (f) Our proposed approach w/ TV (g) full-view Ground Truth (GT) on the $107$th profile of the last subject in the SPARE dataset. Global image, Zoomed image, and Error map are illustrated.}
\label{fig:2}
\end{figure*}

The comparative study evaluates FDK, SIRT, the original DIP using the 3D U-Net~\cite{cciccek20163d} and the UNETR backbones, and our proposed method with and without TV. The original DIP used only the reconstruction loss in Eq.~\eqref{eq:recon}. We employed the one-sided Wilcoxon signed-rank test to evaluate the statistical significance of the comparative study. Table~\ref{tab:1} shows that our approach significantly ($p$-value $<0.05$) outperformed all other methods measured by PSNR and SSIM in terms of both the SPARE and the Walnut datasets. Figure~\ref{fig:2} visualizes the sagittal-plane images of an individual volume map using the same methods in Table~\ref{tab:1}, all reconstructed by $100$-view projection data from a subject in the SPARE dataset. Our proposed approach shows the enhanced resolution in both the zoomed region and the image-domain error map, compared to the other methods. Moreover, all images shown in Fig.~\ref{fig:2} were further evaluated by a radiologist from Sir Run Run Shaw Hospital, Zhejiang University College of Medicine. The report demonstrates that our result presents as relatively clear anatomical structure of the main bronchus as that in the GT image. The images reconstructed by FDK, U-Net and UNETR show ambiguous bronchus due to background noise and low contrast. The reconstructed image produced by SIRT demonstrated severe distortion and spike artifact.

\begin{table}[hb]
\caption{Quantitative results for the reconstruction models of FDK, SIRT, and the original DIP using the 3D U-Net and the UNETR backbones, and our approach (Ours). The mean and the standard deviation (std) of PSNR and SSIM for each method were computed. Bold font highlights the best mean. Regularization parameter $\beta$ was set to $1$ for SPARE while $0$ for Walnut. The asterisk $^*$ denotes the statistical significance ($p$-value$<0.05$) compared to the rest methods in the sense of the one-sided Wilcoxon signed-rank test.} \label{tab:1}
\centering
\scalebox{0.73}{
\begin{tabular}{|c|c|c|c|c|}
\hline
Dataset & \multicolumn{2}{c|}{SPARE} & \multicolumn{2}{c|}{Walnut} \\
\hline
Metric & PSNR (dB) & SSIM & PSNR (dB) & SSIM \\
\hline
FDK     &  $31.15  \pm 0.31$ & $0.929 \pm 0.010$  & $40.74 \pm 1.08$ & $0.977 \pm 0.008$\\
\hline
SIRT    &  $22.12  \pm 0.69$ & $0.753 \pm 0.025$  & $30.75 \pm 1.97$ & $0.873 \pm 0.031$\\
\hline
3D U-Net   &  $31.34  \pm 0.32$ & $0.931 \pm 0.010$  & $41.41 \pm 1.19$ & $0.979 \pm 0.008$   \\
\hline
UNETR   &  $32.85 \pm 0.93$ & $0.947 \pm 0.018$ & $40.91 \pm 0.91$ & $\textbf{0.982} \pm 0.005$  \\
\hline
Ours    &  $\textbf{36.47} \pm 0.58^*$  & $\textbf{0.977} \pm 0.005^*$ & $\textbf{43.02} \pm 1.30^*$ & $\textbf{0.982} \pm 0.007$\\
\hline
\end{tabular}
}
\end{table}



\begin{figure*}[ht]
\centering
\includegraphics[width=\textwidth]{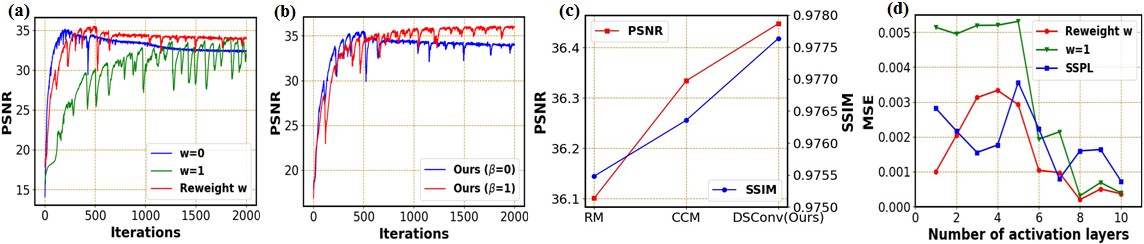}
\caption{Ablation studies for our proposed method. (a) Three different weighting strategies for MSPL with $\beta=0$: w/o MSPL ($\mathbf{w}=0$), fixing weights ($\mathbf{w}=1$), and automatic reweighting strategy (Reweight $\mathbf{w}$); (b) Convergence analysis w/ and w/o the TV penalty by controlling $\beta=1$ and $0$ while using Reweight $\mathbf{w}$; (c) Performance analysis (evaluated by PSNR and SSIM) for three downsampling operations: Resampling method (RM), center-clipping method (CCM), and DSConv; (d) Evaluation of representative features corresponding to activation layers by means of SSPL, and MSPL with $\mathbf{w}=1$ and Reweight $\mathbf{w}$.}
\label{fig:3}
\end{figure*}

We further conducted multiple ablation studies for our proposed method on the SPARE dataset: Figure~\ref{fig:3}(a) compares the switching-off MSPL (through $\mathbf{w}=0$), fixing weights ($\mathbf{w}=1$), and automatically reweighting strategy (Reweight $\mathbf{w}$) for MSPL. We observed that in either strategy, there existed several "dips" on the PSNR curve that frequently caused rapid performance degradation during iteration; also observed in \cite{Chen2021}. The performance of the $\mathbf{w}=0$ strategy declined at the early stage of iteration due to overfitting. Although both the $\mathbf{w}=1$ and the Reweight $\mathbf{w}$ strategies showed relatively consistent increase of performance, the reweighting mechanism obtained fewer 'dips' at the end stage of iteration. In Fig.~\ref{fig:3}(b), we furthermore kept the reweighting mechanism unchanged and evaluated the effectiveness of the TV penalty by controlling $\beta$ as $0$ or $1$. It shows that using the TV penalty ($\beta=1$) preserves the increasing of performance compared to that without the TV penalty, which ensures our proposed method to optimize continually and converge without any additional requirement such as early stopping. In Fig.~\ref{fig:3}(c), we show performance of other downsampling operations for DSConv, namely, the resampling method (RM) using spline interpolation and the center-clipping method (CCM) by selecting the middle-three consecutive slices. The learnable DSConv obtained the best output image quality than the other non-learning downsampling operations. Lastly, we evaluated how representative the perceptual features were at different resolution scales for MSPL. We calculated all VGG-11 activation inputted by both the FDK reference image and the generated LVM output image, and then obtained MSE between the corresponding activation features related to the two images. We substituted MSPL with single scale PLs (SSPL) which obtains perceptual features only on a single VGG-11 activation layer, and apply fixing-weight MSPL ($\mathbf{w}=1$) and reweight $\mathbf{w}$ on the first $n$ activation layers for comparison. Figure~\ref{fig:3}(d) indicates that the predictive features by MSPL (Reweight $\mathbf{w}$) obtained the lowest error compared to the others when $n\geq 5$.

\section{Discussion and Conclusion}
\label{sec:discon}

Our proposed method is a novel forward-model-free LVM-based DIP framework with MSPL for the sparse-view 3D CBCT reconstruction using the adaptively reweighting one-shot optimization. The quantitative and qualitative evaluations demonstrate that the proposed approach effectively enhances the reconstructed image quality and ensures the convergence to the full-view GT image. Our approach can potentially extend to other tomographic imaging modalities. In the future, we will investigate broadening generalization bound of the DIP-type method suggested in ~\cite{Zou2021} and evaluate the clinical viability of our approach.

\newpage

\bibliographystyle{IEEEbib}
\bibliography{strings,ref2}

\end{document}